\documentclass[12pt]{iopart}

\usepackage[pdftex]{graphicx}
\usepackage{iopams}
\usepackage{epstopdf}
\usepackage{graphicx}
\usepackage{cite}
\usepackage{gensymb}

\begin{document}
\def\be{\begin{equation}}
\def\ee{\end{equation}}
\def\bea{\begin{eqnarray}}
\def\eea{\end{eqnarray}}
\def\rp{r_{+}}
\def\rmm{r_{-}}

\title{
\bf A Gauge Invariant Dual Gonihedric 3D Ising Model
}
\date{May 2011}
\author{D. A. Johnston}
\address{Dept. of Mathematics, Heriot-Watt University,
Riccarton, Edinburgh, EH14 4AS, Scotland}

\author{R. P. K. C. M. Ranasinghe}
\address{Department of Mathematics, University of Sri Jayewardenepura,
Gangodawila, Sri Lanka.}


\begin{abstract}
We note that two formulations of dual gonihedric Ising models in $3d$, one based 
on using Wegner's general framework for duality to construct a dual Hamiltonian for codimension one surfaces, 
the other on constructing a dual Hamiltonian for two-dimensional surfaces, are related 
by a variant of the standard decoration/iteration transformation. 

The dual Hamiltonian for two-dimensional surfaces contains a mixture of link and vertex spins
and as a consequence possesses a gauge invariance which is inherited by the codimension one
surface Hamiltonian. This gauge invariance  ensures  
the latter is equivalent to a third formulation, an anisotropic Ashkin-Teller model. We describe
the equivalences in detail and discuss
some Monte-Carlo simulations which support these observations.

\end{abstract} 

\maketitle


\section{Introduction}

The dual of the standard Ising Hamiltonian with nearest neighbour $\langle ij \rangle$ couplings
on a $3d$ cubic lattice
\begin{equation}
\label{e0I}
H_{Ising} =  -  \sum_{\langle ij \rangle}\sigma_{i} \sigma_{j}
\end{equation}
is the $\mathbb{Z}_2$ Ising gauge theory
\begin{equation}
\label{e0G}
H_{Gauge} =  -  \sum_{[ijkl]} U_{ij} U_{jk} U_{kl} U_{li}
\end{equation}
where the sum is over plaquettes $[ijkl]$ and the spins live on the edges of the lattice. The coupling $\beta$ in the partition function $Z(\beta) = \sum_{\{ \sigma \}} \exp ( - \beta H_{Ising})$
and its dual $\beta^*$ 
in $Z(\beta^*) = \sum_{\{ U \}} \exp ( - \beta^* H_{Gauge})$
are related by $\beta^* = -(1/2) \log \, \tanh  \beta  $.

In this paper we will investigate the relation between three (apparently) different
formulations of the dual to the {\it gonihedric} Ising model  \cite{1} 
\begin{equation}
\label{e2k}
H_{\kappa=0} =  -  \sum_{[ijkl]}\sigma_{i} \sigma_{j}\sigma_{k} \sigma_{l}
\end{equation}
which, like the Ising gauge theory, has a plaquette interaction but in which the spins now reside
at the vertices of the $3d$ lattice. The subscript $\kappa=0$ appears because this plaquette
Hamiltonian is a particular case of a one-parameter family of gonihedric Hamiltonians
\footnote{We have dropped a factor of $1/2$ in the coupling definition compared with \cite{5,5b} in order
to keep the standard definition of the duality relations here.}
\begin{equation}
\label{e1}
H_{gonihedric} = - 4 \kappa \sum_{\langle ij\rangle }\sigma_{i} \sigma_{j}  +
\kappa \sum_{\langle \langle ij\rangle \rangle }\sigma_{i} \sigma_{j} 
- ( 1-\kappa )\sum_{[ijkl]}\sigma_{i} \sigma_{j}\sigma_{k} \sigma_{l} \; .
\end{equation}
defined by Savvidy and Wegner \cite{2}, where the $\langle \langle ij\rangle \rangle$
are next-to-nearest neighbour sums.
The spin cluster boundaries of this Hamiltonian were intended to mimic
a gas of worldsheets arising from a gonihedric string action. 
When discretized using triangulations, this action may be written as 
\begin{equation}
S = {1 \over 2} \sum_{\langle ij \rangle} | \vec X_i - \vec X_j | \; \theta (\alpha_{ij}),
\label{steiner}
\end{equation}
where
$\theta(\alpha_{ij}) = | \pi - \alpha_{ij} |$, $\alpha_{ij}$ is the dihedral angle between the
neighbouring triangles with a common edge $\langle ij \rangle$
and  $| \vec X_i - \vec X_j |$ are the lengths of the triangle edges. 

The word gonihedric was originally coined to reflect the properties of this action
which weights edge lengths  between non-coplanar triangles rather than their areas.
It combines the Greek words gonia
for angle, referring to the dihedral angle, and hedra for base or face, referring
to the adjacent triangles. $H_{gonihedric}$ is an appropriate cubic lattice discretization 
of such an action because it too assigns zero weight to the areas of spin cluster
boundaries, rather weighting edges and intersections \cite{3}. This gives $H_{gonihedric}$ very different properties
to $H_{Ising}$ where (only) the areas of spin cluster boundaries are weighted.

The plaquette action $H_{\kappa=0}$  has been shown to possess 
a  degenerate low-temperature phase and a first order phase transition as well as interesting,
possibly glassy,  dynamical properties \cite{4}. It displays a peculiar ``semi-global'' symmetry in which planes of spins may be flipped at zero energy cost, accounting for the degeneracy of the low temperature phase. For non-zero $\kappa$ this symmetry appears to be broken at finite temperature
and the transition becomes second order. In \cite{5} we observed that one formulation of the dual to $H_{\kappa=0}$, which took the form of an anisotropic Ashkin-Teller model, displayed similar symmetry properties since it was possible to flip planes of spins in this also.
In \cite{5b} we related this to a superficially different dual formulation that employed three flavours of spins by using a gauge-fixing procedure. There were indications of potentially interesting dynamical behaviour
for both Hamiltonians in \cite{5} and \cite{5b}.

In the following we discuss the derivation of these
two dual Hamiltonians as well as a {\it third} possibility. We then describe the relation 
between the various Hamiltonians via a decoration transformation and gauge-fixing, before outlining some Monte-Carlo simulations in support of these observations.

\section{Duals Galore} 

The dual to $H_{\kappa=0}$ was initially constructed ``by hand'' by Savvidy {\it et.al.} \cite{6} 
by considering the high temperature expansion of the plaquette Hamiltonian 
\begin{eqnarray}
Z (\beta)  &=& \sum_{\{\sigma\}}  \exp (- \beta H_{\kappa=0} ) \nonumber \\
&=& \sum_{\{\sigma\}} \prod_{[ijkl]} \cosh  \left(\beta \right) \left[1 + \tanh \left(\beta \right) ( \sigma_i \sigma_j \sigma_k \sigma_l )\right] 
\label{z2}
\end{eqnarray}
which can be written as
\begin{equation}
Z (\beta)  = \left[ 2 \cosh \left(\beta \right) \right]^{3 L^3} \sum_{\{ S \}} \left[\tanh \left(\beta \right)\right]^{n ( S ) } 
\label{z2a}
\end{equation}
on an $L^3$ cubic lattice, where the sum runs over closed surfaces with an even number of plaquettes at any vertex. In the summation $ n ( S )$ is the number of 
plaquettes in a given surface. 

Surprisingly, the low temperature expansion (i.e. high temperature in the dual variable $\beta^* = - (1/2)  \log \, \tanh \beta$)
of the following
{\it anisotropic} Hamiltonian 
\begin{equation}
\label{dual0}
 H_{dual0} = - \sum_{\langle ij \rangle} \sigma_{i}  \sigma_{j} 
- \sum_{\langle ik \rangle } \tau_{i}  \tau_{k} 
-  \sum_{ \langle jk \rangle} \eta_{j} \eta_{k} 
\end{equation}
produced the requisite diagrams. In $H_{dual0} $ the sums are one-dimensional and run along
the  orthogonal axes, with $ij, ik$ and $jk$ representing the
$z$, $y$ and $x$ axes respectively using our conventions. 
The spins are non-standard and live in the fourth order Abelian group, since
the geometric constraint on the plaquettes means that
\begin{eqnarray}
e \sigma &=&  \sigma \; , \;  \; e \tau = \tau \; ,  \; \; e \eta = \eta \nonumber \\	
\sigma^2 &=& \tau^2 = \eta^2 = e \\
\sigma \tau &=& \eta \, , \; \; \tau \eta = \sigma \; , \; \; \eta \sigma = \tau \nonumber
\end{eqnarray}
with $e$ being the identity element. They can be thought of as representing differently oriented
matchbox surfaces such as that shown in Fig.~(1), which are combined by facewise multiplication. The shaded
faces carry a negative sign and the associated spin variable lives at the centre
of the matchbox. Any spin cluster boundary in the model can be constructed from such matchboxes while still satisfying the local constraint on the number of incident plaquettes.
\begin{figure}[h]
\begin{center}
\includegraphics[height=4cm]{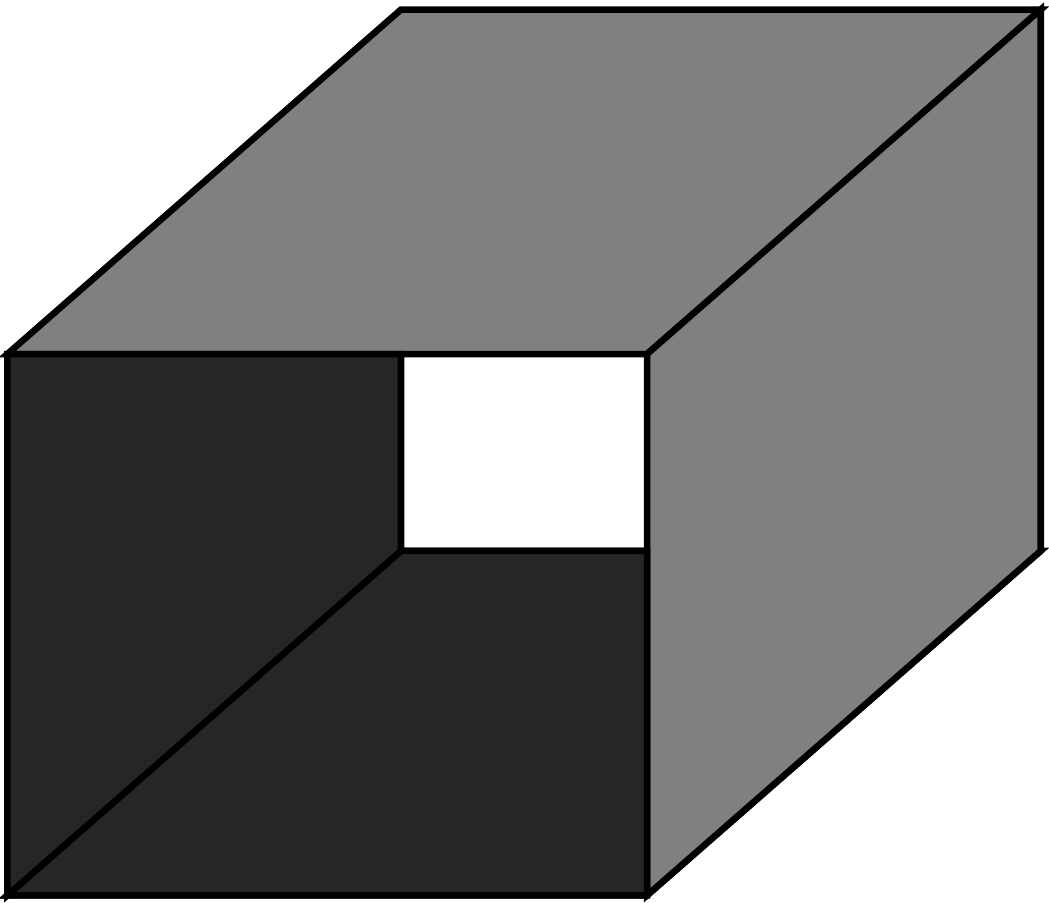}
\label{matchbox} 
\caption{An elementary matchbox surface represented by one of the spins in equ.~(\ref{dual0})}
\end{center}
\end{figure}

The spins may also be taken to be Ising ($\pm 1$) variables
if we set $\eta_i = \sigma_i \; \tau_i $, which is more convenient for simulations.
This modifies $H_{dual0}$ to an anisotropically coupled Ashkin-Teller Hamiltonian \cite{AT}
\begin{equation} 
\label{dual1}
H_{dual1} = -  \sum_{ \langle ij \rangle} \sigma_{i}  \sigma_{j} 
-  \sum_{ \langle ik \rangle } \tau_{i}  \tau_{k} 
-   \sum_{\langle jk \rangle} \sigma_{j} \sigma_{k} \tau_{j}  \tau_{k} \, . 
\end{equation}
We investigated this formulation of the dual model in \cite{5} and found that it
displayed a first order phase transition and similar semi-global symmetries to those of $H_{\kappa=0}$. The symmetries
were a direct consequence of the anisotropic couplings, which allowed a greater freedom in transforming the spin variables than in the isotropically coupled version of equ.~(\ref{dual1}), which is just the Ashkin-Teller model at its four-state Potts point.

It is also possible to construct duals to $H_{\kappa=0}$ and its higher dimensional equivalents \cite{6b} using the completely general framework
for duality in lattice spin models that was
originally formulated by Wegner in \cite{6c}. There are two possible
ways to write the dual to $H_{\kappa=0}$ in three dimensions with this machinery, using either
the general formula for the dual of codimension one surfaces or
the  formula for the dual of two dimensional surfaces in $d$ dimensions. 
If we temporarily use the notation of \cite{6b},
the dual Hamiltonian
for a codimension one surface in $d$ dimensions is given by 
\bea
\label{codim1}
H^{d}_{dual,  codim 1} &=& - \sum_{\alpha<\beta, \,  \vec r}\prod_{\gamma}
\Lambda_{\alpha,\beta\gamma}(\vec r)
\Lambda_{\alpha,\beta\gamma}(\vec r+\vec e_{\gamma})
\Lambda_{\beta,\alpha\gamma}(\vec r)
\Lambda_{\beta,\alpha\gamma}(\vec r+\vec e_{\gamma}) \nonumber \\
{}
\eea
where the $\Lambda$ spins live on each of the $(d-3)$ dimensional (hyper)vertices 
situated at the vertices $\vec r$ of the hypercubic lattice and the indices $\alpha, \beta,\gamma$
run from $1$ to $d$. The unit vectors  $\vec e_{\gamma}$ point along the lattice axes. 
The
dual Hamiltonian for a two-dimensional gonihedric surface embedded in $d$ dimensions is of the form 
\be
\label{d2d}
H^{d}_{dual, \, 2d} = -\sum_{\vec r}\sum_{\beta \neq \gamma}
\Lambda_{\beta\gamma}(\vec r)
\Gamma(\vec r,\vec r +\vec e_{\gamma})
\Lambda_{\beta\gamma}(\vec r+\vec e_\gamma)
\ee
where we now have $\Gamma$ spins on each (hyper)edge
in addition to the $\Lambda$ spins at each vertex.

If we  specialize to two dimensional surfaces embedded in three dimensions, which is the case
for the dual of $H_{\kappa=0}$, either formulation may be employed. 
Returning to our own notation \cite{5b}, the codimension one Hamiltonian of equ.~(\ref{codim1}) in three dimensions may be written as 
\begin{equation}
\label{dual3a}
H_{dual2} = - \sum_{\langle ij \rangle} \sigma_i \sigma_j \mu_i \mu_j
- \sum_{\langle ik \rangle } \tau_i \tau_k  \mu_i \mu_k
- \sum_{ \langle jk \rangle} \sigma_j \sigma_k \tau_j \tau_k \, ,
\end{equation}
where we again have one-dimensional sums as with $H_{dual0}$ and $H_{dual1}$, but there are now three
flavours of spins living at each vertex which display a local Ising gauge symmetry
$
\sigma_i, \tau_i, \mu_i \to \gamma_i \sigma_i, \gamma_i \tau_i, \gamma_i \mu_i
$
in addition to the planar flip symmetries of $H_{dual1}$. 

Still within the general approach of Wegner \cite{6c}, in three dimensions the Hamiltonian of equ.~(\ref{d2d}) for the two-dimensional surface variant   also contains three flavours of vertex spins
$\sigma_i, \tau_i , \mu_i$, but in addition there are spin variables $U_{ij}^{1,2,3}$ living on the  lattice edges 
which couple in an anisotropic manner to the vertex spins 
\bea
\label{sUs}
H_{dual3} &=& - \sum_{\langle ij \rangle} \left( \sigma_i U^{1}_{ij} \sigma_j  + \mu_i U^{1}_{ij} \mu_j \right) 
- \sum_{\langle ik \rangle } \left( \tau_i U^{2}_{ik} \tau_k +  \mu_i  U^{2}_{ik} \mu_k \right) \nonumber \\
&-&  \sum_{ \langle jk \rangle} \left( \sigma_j U^{3}_{jk} \sigma_k  + \tau_j U^{3}_{jk} \tau_k  \right) \; .
\eea 
We thus have three superficially rather different Hamiltonian formulations for the, presumably 
unique, dual of the plaquette Hamiltonian $H_{\kappa=0}$ in three dimensions: 
\begin{itemize}
\item{} $H_{dual3}$ in equ.~(\ref{sUs}) containing both vertex and edge spins
\item{} $H_{dual2}$ in equ.~(\ref{dual3a}) containing purely four spin interactions
\item{} $H_{dual1}$ in equ.~(\ref{dual1}) which is Ashkin-Teller in form.	
\end{itemize} 
In the the next section we discuss
the relation between $H_{dual3}$ and $H_{dual2}$, and thereafter that between $H_{dual2}$
and $H_{dual1}$.

\section{Decoration}

The equivalence between $H_{dual3}$ and $H_{dual2}$  is a
consequence of a variation of the classical decoration transformation \cite{6b,7}. In the standard transformation
an edge with spins $\sigma_1, \sigma_2$ at each vertex is decorated with a link spin $s$
as in Fig.~(2).
\begin{figure}[h]
\label{fig0a}
\centering
\includegraphics[height=4cm]{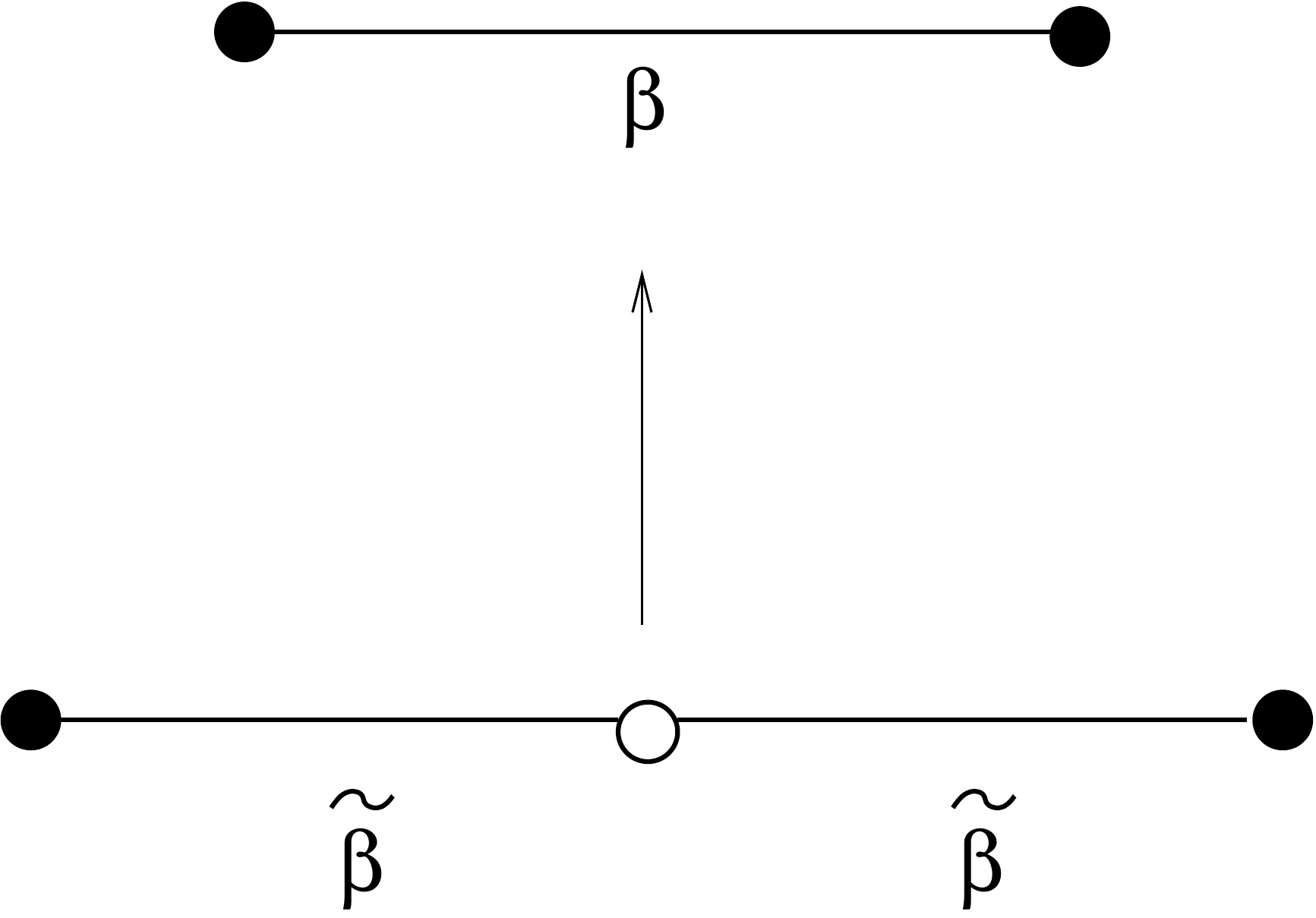}
\caption{The standard decoration transformation}
\end{figure}
If the coupling between $s$ and $\sigma_{1}$ and $\sigma_{2}$ is $\tilde \beta$, 
summing over the central spin $s$ gives rise to a 
new effective coupling $\beta$ between the 
primary vertex spins $\sigma_1, \sigma_2$
\begin{equation}
\label{decor}
\sum_{s} \exp \left[ \tilde \beta s ( \sigma_1 + \sigma_2 ) \right] = A \, \exp ( \beta \sigma_1 \sigma_2).
\end{equation}
Both the prefactor $A$ and the coupling $\beta$ may be expressed in terms of $\tilde \beta$
by enumerating possible spin configurations in equ. (\ref{decor}).
This gives 
\bea
A &=& 2  \cosh ( 2 \tilde \beta )^{1/2} \nonumber \\
\beta &=&   \frac{1}{2}  \log  \cosh ( 2 \tilde \beta ) .
\label{e2}
\eea
We can repeat this procedure with the $U$ spins on each edge in $H_{dual3}$. In this case
each direction has two flavours of vertex spin and performing the sum generates
the four-spin couplings of $H_{dual2}$, for example
\bea
\sum_{\{U^1_{12}\}} \exp \left[ \tilde \beta   \left( \sigma_1 U^1_{12} \sigma_2  + \mu_1 U^1_{12} \mu_2 \right) \right]
= A  \, \exp ( \beta \sigma_1 \sigma_2 \mu_1  \mu_2) .
\eea
The sum over $U$ may be carried out globally over every edge which immediately demonstrates
equivalence of the partition functions for $H_{dual3}$ and $H_{dual2}$
\bea
\label{sUs2}
Z &=& \sum_{\{U, \sigma\}} \exp [ -\tilde \beta H_{dual3} ] \nonumber \\
 &=& \sum_{\{U, \sigma\}} \exp \left [ \tilde \beta\sum_{\langle ij \rangle} \left( \sigma_i U^{1}_{ij} \sigma_j  + \mu_i U^{1}_{ij} \mu_j \right) 
+ \tilde \beta \sum_{\langle ik \rangle } \left( \tau_i U^{2}_{ik} \tau_k +  \mu_i  U^{2}_{ik} \mu_k \right) \nonumber \right. \\
&+& \left. \tilde \beta \sum_{ \langle jk \rangle} \left( \sigma_j U^{3}_{jk} \sigma_k  + \tau_j U^{3}_{jk} \tau_k  \right) \right] \\
&=& B \, \sum_{\{\sigma\}}  \exp \left[  \beta \left( \sum_{\langle ij \rangle} \sigma_i \sigma_j \mu_i \mu_j + \sum_{\langle ik \rangle } \tau_i \tau_k  \mu_i   \mu_k  +  \sum_{ \langle jk \rangle}  \sigma_j  \sigma_k   \tau_j \tau_k \right)  \right] \nonumber \\
&=& B \, \sum_{\{\sigma\}} \exp [ -\beta H_{dual2} ] \nonumber .
\eea 
The overall factor
$B$ coming from a product of $A$'s on the individual links
is irrelevant for calculating physical quantities and the two couplings are again related by the standard
decoration relation, $\beta =   (1/2)  \log  \cosh ( 2 \tilde \beta )$.

\section{Gauge Fixing (and Flips)}

The equivalence between $H_{dual2}$ and $H_{dual1}$, on the other hand, is a consequence of the 
additional
gauge symmetry \cite{5b} which is present in $H_{dual2}$
\be
\sigma_i, \, \tau_i, \, \mu_i \, \to \,  \gamma_i  \sigma_i, \,  \gamma_i \tau_i, \, \gamma_i \mu_i \, .
\ee
We are at liberty to choose the Ising spin gauge transformation parameter $\gamma_i$ to be equal to 
one of the spin values, say $\mu_i$, at each site
so the gauge transformation then becomes
\be
\sigma_i, \, \tau_i, \, \mu_i \, \to \,  \mu_i  \sigma_i, \,  \mu_i \tau_i, \, 1
\ee
which, using the fact that the sum over the remaining spin variables $\sigma_i, \tau_i$ is invariant under the transformation, relates the partition functions for the two Hamiltonians as
\bea 
Z &=& \sum_{ \{\sigma,\tau,\mu \}} \exp \left[ - \beta H_{dual2} ( \sigma,\tau,\mu) \right] \nonumber \\
&=& 2^{L^3} \sum_{\{ \sigma,\tau \}} \exp \left[ - \beta H_{dual2} ( \sigma,\tau,\mu=1) \right] \\
&=&  2^{L^3} \sum_{\{\sigma,\tau \}} \exp \left[ - \beta H_{dual1} ( \sigma,\tau) \right] \; .
\nonumber
\eea
The coupling $\beta$ is not transformed in this case and we can, of course, choose to eliminate any one of the 
three spins, which simply amounts to relabelling the axes. From this perspective $H_{dual1}$ is simply a gauge-fixed
version of $H_{dual2}$. This can be  confirmed by Monte-Carlo simulations which measure the same energies (and energy distributions)
and transition points for the observed first order phase transitions \cite{5b}.

The equivalence between $H_{dual3}$ and $H_{dual2}$ described in the preceding section via the decoration transformation also sheds light on the somewhat unexpected presence of this gauge symmetry in  $H_{dual2}$. All the terms in $H_{dual3}$ are of the gauge-matter coupling form $ \sigma_i U_{ij} \sigma_j$, so this action possesses a similar, standard gauge invariance to that seen in other gauge-matter systems such as 
the $\mathbb{Z}_2$ gauge-Higgs model, namely
\bea    
\sigma_i \to \gamma_i \sigma_i \; , \; \sigma_j \to \gamma_j \sigma_j \; , \; U_{ij}^{1,3} \to \gamma_i U_{ij}^{1,3} \gamma_j \nonumber \\
\tau_i \to \gamma_i \tau_i \; , \; \tau_j \to \gamma_j \tau_j \; , \; U_{ij}^{2,3} \to \gamma_i U_{ij}^{2,3} \gamma_j \\
\mu_i \to \gamma_i \mu_i \; , \; \mu_j \to \gamma_j \mu_j \; , \; U_{ij}^{1,2} \to \gamma_i U_{ij}^{1,2} \gamma_j \nonumber \; .
\eea
When the $U$ spins are summed over to give  $H_{dual2}$, the gauge symmetry of the $\sigma, \tau$ and $\mu$
spins remains as an echo of this symmetry. In both cases if we look
at a single site transformation all three spins $\sigma_i, \tau_i$ and $\mu_i$ must be transformed. In  $H_{dual3}$ this is a consequence of the way in which the three edge spins $U^{1,2,3}_{ij}$ couple to the vertex spins.

A characteristic feature of both $H_{dual2}$ and $H_{dual1}$ is the flip symmetry of the low temperature phase, which allows planes of spins to be flipped at zero energy cost.
This can be observed in the ground state by decomposing the full lattice Hamiltonian into cube terms and searching for minimum energy configurations on a single cube. The absence of any sign of non-zero magnetic order parameters in the low temperature phase in Monte-Carlo simulations then indicates
that this ground-state symmetry persists throughout the low temperature phase. 
More specifically, for  $H_{dual2}$ it is possible to flip planes of  pairs of spins at no energy cost. Similarly, planes of either one or two spins 
depending on the orientation may  be flipped in $H_{dual1}$.
Similar behaviour is also seen with the original $H_{\kappa=0}$ plaquette Hamiltonian, where the symmetry has been confirmed to persist into the low-temperature phase in low-temperature expansions by Pietig and Wegner \cite{10}.

In $H_{dual3}$ 
flipping the three spins at a vertex may be compensated  locally by flipping the six incident edge spins as shown in Fig.~(3), which is  the local gauge transformation that is still present in $H_{dual2}$.
\begin{figure}[h]
\begin{center}
\includegraphics[height=6cm]{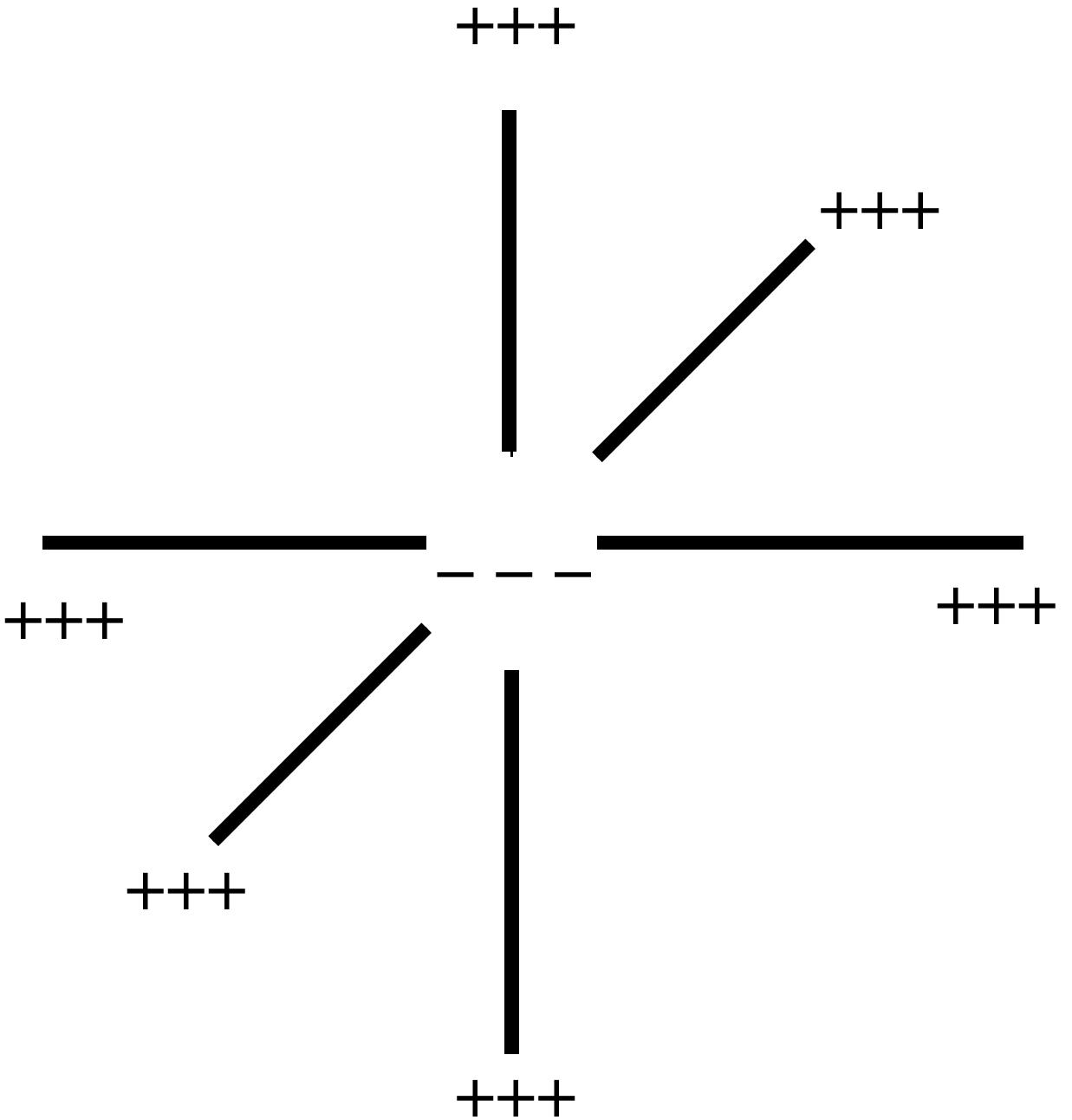}
\label{matchbox1} 
\caption{Flipping three spins at a site may be compensated by flipping the six incident edge spins shown in bold, a purely local gauge transformation.}
\end{center}
\end{figure}
On the other hand, if just two spins are flipped at the central site, e.g. $\sigma$ and $\mu$, the gauge transformation can no longer be applied to keep the disturbance local. However a global, planar spin flip can still leave the energy unchanged, as shown in Fig.~(4).  
Choosing to flip different pairs of  spins at the central vertex 
can be compensated  by flipping the appropriate pairs of incident edge spins ($U^2$ or $U^3$) and  differently oriented planes of vertex spins. The planar flip symmetry is thus still a feature of the ground state of $H_{dual3}$ and distinct from the gauge symmetry.
\begin{figure}[h]
\begin{center}
\includegraphics[height=6cm]{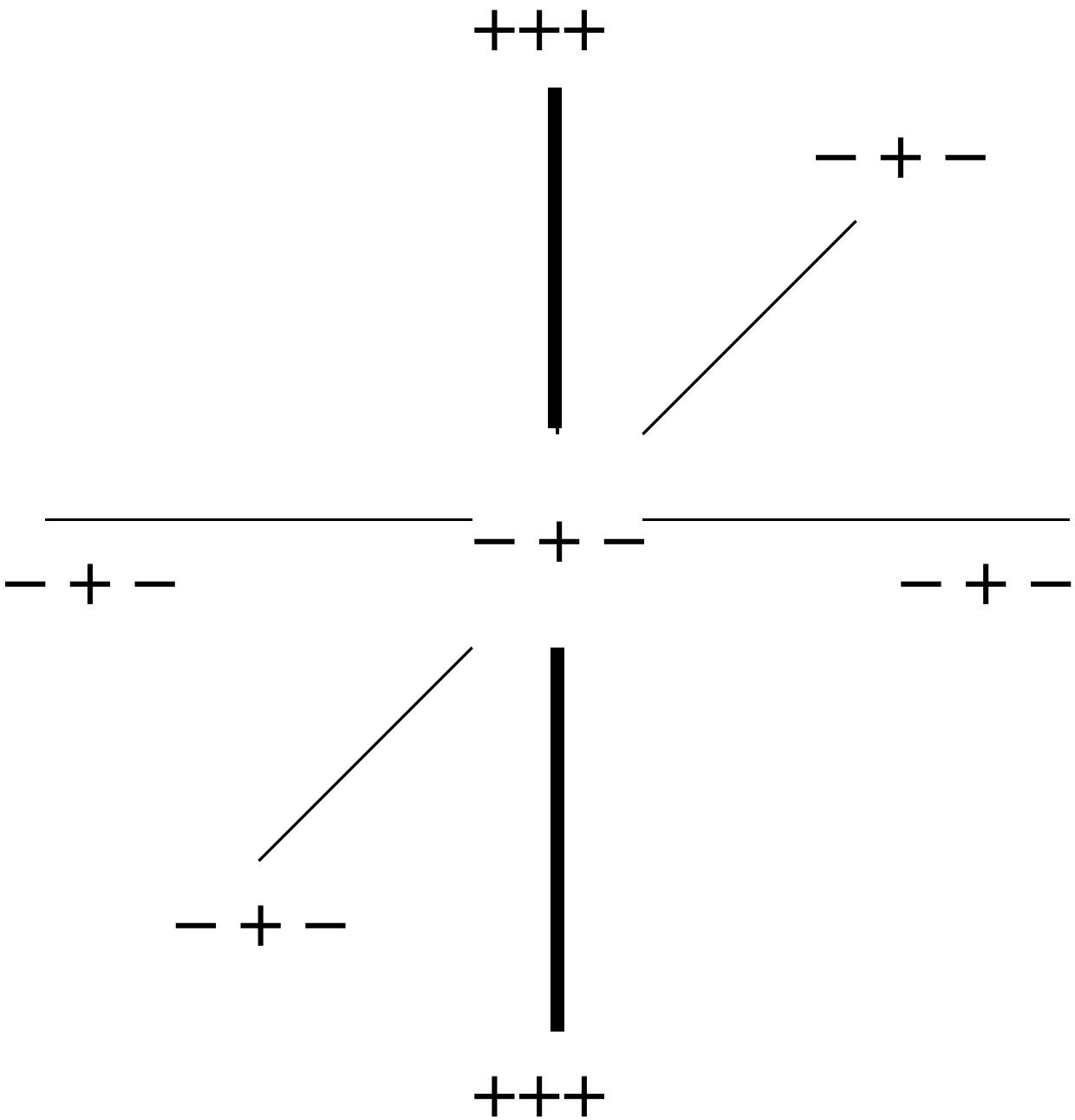}
\label{matchbox2} 
\caption{Flipping two spins ($\sigma, \mu$ in this case) at a site may be compensated by flipping two incident edge spins ($U^1$) shown in bold along with the other coplanar vertex spins. This is no longer a purely local gauge transformation since the motif must be propagated across the lattice to maintain the same energy.}
\end{center}
\end{figure}

\section{Monte Carlo}
 
\begin{figure}[h]
\centering
\includegraphics[height=6cm]{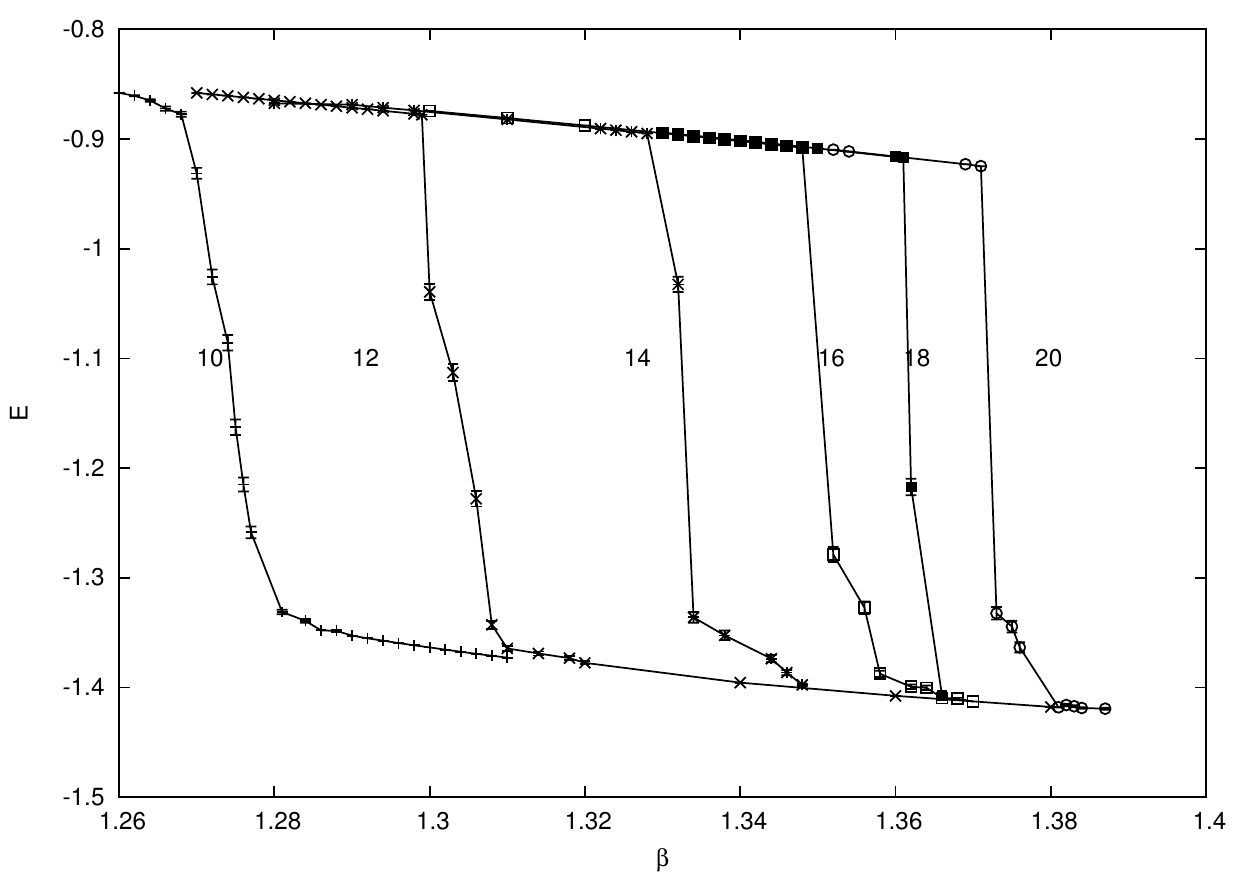}
\caption{The energy for $H_{dual2}$ on lattices ranging from $10^3$ to $20^3$ from left to right. The lines joining the data points are drawn to guide the eye. Data from $H_{dual1}$ is essentially identical.}
\label{E02} 
\end{figure}

Monte-Carlo simulations reveal a first order phase transition in both $H_{dual2}$ and $H_{dual1}$ 
at $\beta \simeq 1.39$ \cite{5,5b}. 
This can be seen in measurements of the energy, where there is a sharp drop
at the transition point. A plot of the energy is shown for various lattice sizes 
in Fig.~(\ref{E02}) for $H_{dual2}$, the values for $H_{dual1}$ are essentially identical. 
The first order nature of the transition for  $H_{dual2}$ and $H_{dual1}$  is confirmed by 
observing a dual peak structure in the energy histogram $P(E)$ near the transition point and a non-trivial value  of Binder's energy cumulant
\begin{equation}
	U_E = 1 - \frac{\langle E^4 \rangle}{3  \langle E^2 \rangle^2} 
\end{equation}
as a consequence of the shape of $P(E)$.

Based on these observations, and allowing for a factor of $1/2$ in our definitions of $H_{dual1}$ and $H_{dual2}$ in \cite{5,5b}, we would expect to see
a transition in $H_{dual3}$ at the the value of $\tilde \beta$ 
found by inverting the decoration transformation, namely $(1/2) \cdot \cosh^{-1}(\exp(1.39))=1.034$
in the thermodynamic limit. 
To confirm this expectation, we carried out Monte-Carlo simulations using $10^3, 12^3, 16^3$ and $18^3$  lattices with periodic boundary conditions for all spins 
 at various temperatures with a simple Metropolis update. After  an appropriate number of thermalization sweeps, $10^7$ measurement sweeps were carried out at each lattice size for each temperature.  
We have not attempted to construct a cluster algorithm since they do not offer effective speedup at first order transition points such as that (presumably) under investigation here and because of the additional complication
of the edge spins.

\begin{figure}[h]
\centering
\includegraphics[height=6cm]{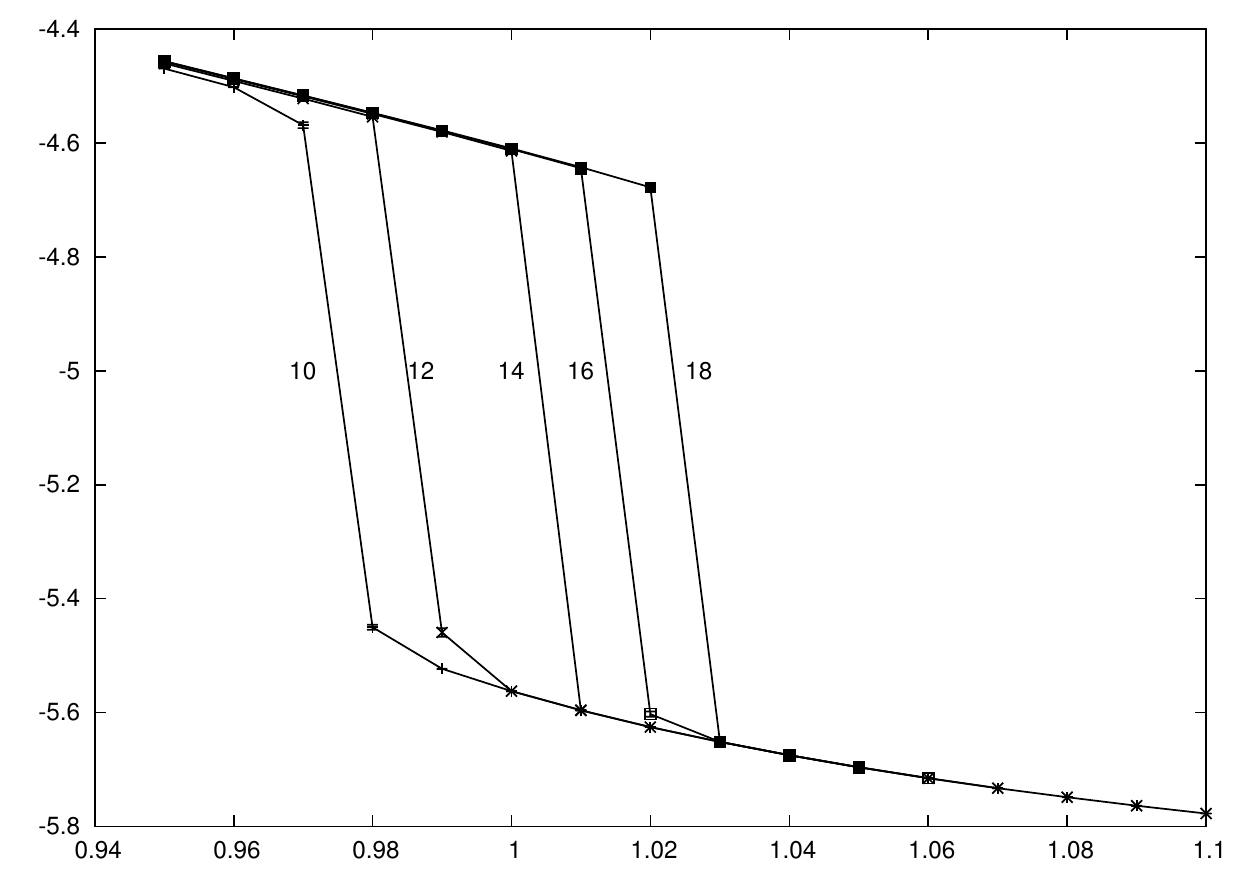}
\caption{The energy for $H_{dual3}$ on lattices ranging from $10^3$ to $18^3$ from left to right. The lines joining the data points are drawn to guide the eye.}
\label{E0c} 
\end{figure}

Looking at measurements of the energy from our simulations of $H_{dual3}$ in Fig.~(\ref{E0c}) we can see that a similar sharp drop in the energy consistent with a 
first order transition is still present.
The observed finite size transition temperatures agree well with those calculated by transforming
the values from Fig.~(4) using the decoration relation, e.g. for $L=10$ we would expect  $\beta_c = 0.5 \times \cosh^{-1}(\exp (1.275)) = 0.975$, as found directly in the simulation.

\begin{figure}[h]
\centering
\includegraphics[height=6cm]{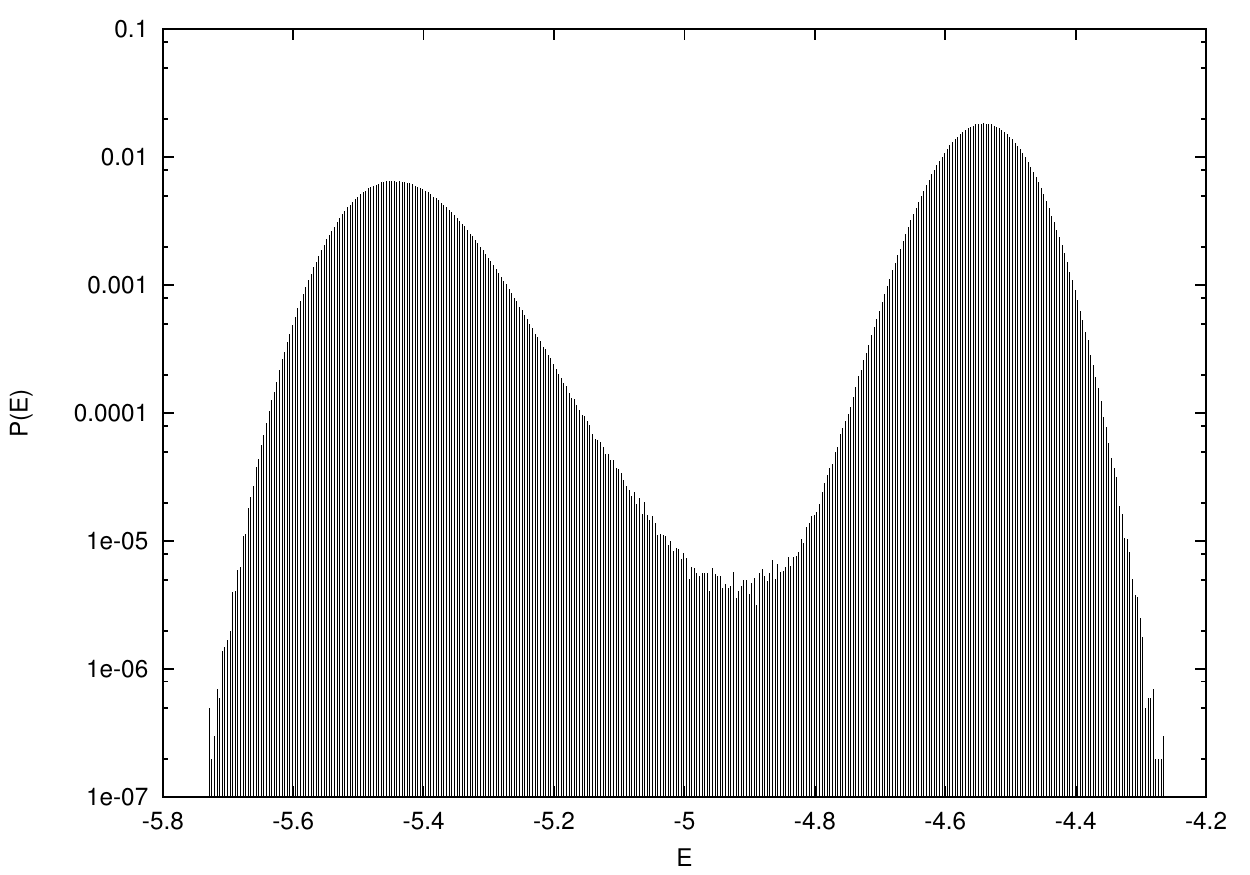}
\caption{The energy histogram $P(E)$ close to the pseudocritical point at $\beta=0.972$ on a $10^3$ lattice.}
\label{PE} 
\end{figure}

Further evidence for a first order transition with $H_{dual3}$, as noted above for the other dual Hamiltonians, can be garnered by looking at the energy histogram $P(E)$ to discern a dual peak structure. In Fig.~(\ref{PE})  close to the pseudocritical point for $L=10$ at $\beta=0.972$ we can see clear evidence of two peaks.
The Monte-Carlo simulations thus lend support to the observation that $H_{dual3}$ and $H_{dual2}$ are related by 
a decoration transformation through the agreement of the suitably transformed transition temperatures and also
confirm the first order nature of the transition seen in $H_{dual3}$.
There is no signal for the transition in the magnetic quantities $\langle \sigma \rangle$, $\langle \tau \rangle$, $\langle \mu \rangle$ due to the gauge invariance and flip symmetries of the Hamiltonian, so in this respect too $H_{dual3}$ is similar to $H_{dual2}$ and $H_{dual1}$.

From the point of view of efficient simulation, an application of a principle of least effort
suggests that the best adapted dual formulation for numerical work is probably that of 
the Ashkin-Teller like Hamiltonian of $H_{dual1}$ in \cite{5}, since that has the minimum number of spin degrees of freedom
to simulate. $H_{dual2}$ adds an additional flavour of vertex spin to this that is effectively a gauge degree of freedom and $H_{dual3}$ uncouples the four spin interactions in this with further
edge spins. The physics of all three dual Hamiltonians is the same. 
  
\section{Discussion}

To summarize, we have  the following chain of equivalences between the various dual gonihedric Hamiltonians
in $3d$
\bea
H_{dual3} &=& - \sum_{\langle ij \rangle} \left( \sigma_i U^{1}_{ij} \sigma_j  + \mu_i U^{1}_{ij} \mu_j \right) 
- \sum_{\langle ik \rangle } \left( \tau_i U^{2}_{ik} \tau_k +  \mu_i  U^{2}_{ik} \mu_k \right)  \nonumber \\
&-& \sum_{ \langle jk \rangle} \left( \sigma_j U^{3}_{jk} \sigma_k  + \tau_j U^{3}_{jk} \tau_k  \right) \nonumber \\
{} \nonumber \\
&{}&\longrightarrow \textrm{(Un)Decoration} \longrightarrow \nonumber \\
{} \nonumber \\
H_{dual2} &=& - \sum_{\langle ij \rangle} \sigma_i \sigma_j \mu_i \mu_j
- \sum_{\langle ik \rangle } \tau_i \tau_k  \mu_i \mu_k 
- \sum_{ \langle jk \rangle} \sigma_j \sigma_k \tau_j \tau_k \nonumber \\
{} \nonumber \\
&{}& \longrightarrow 
\textrm{Gauge-Fixing} \longrightarrow
 \\
{} \nonumber \\
H_{dual1} &=& - \sum_{ \langle ij \rangle} \sigma_{i}  \sigma_{j} 
- \sum_{ \langle ik \rangle } \tau_{i}  \tau_{k} 
-  \sum_{\langle jk \rangle} \sigma_{j} \sigma_{k} \tau_{j}  \tau_{k} \nonumber \\
{} \nonumber \\
&{}& \longrightarrow 
\textrm{Non-Ising variables} \longrightarrow \nonumber \\
{} \nonumber \\
H_{dual0} &=& - \sum_{\langle ij \rangle} \sigma_{i}  \sigma_{j} 
- \sum_{\langle ik \rangle } \tau_{i}  \tau_{k} 
-  \sum_{ \langle jk \rangle} \eta_{j} \eta_{k} \nonumber 
\eea
In the above  we have listed the operations relating the various formulations 
in three dimensions. A variant of the classical decoration transformation in which edge spins are summed out relates $H_{dual3}$ to $H_{dual2}$. 
In transforming $H_{dual3} \to H_{dual2}$ the coupling is therefore transformed as $\beta =  (1 / 2) \ln \, \cosh ( 2 \tilde \beta )$.
The gauge-invariant nature of $H_{dual3}$ due to the presence of both edge and vertex spins
leaves an echo in the vertex spin gauge symmetry of $H_{dual2}$, which in turn ensures the equivalence of $H_{dual2}$ and $H_{dual1}$ via
a gauge-fixing. Allowing non-Ising spins gives a final equivalence between the dual models
$H_{dual1}$ and $H_{dual0}$ and a standard duality transformation then takes us back to the original plaquette gonihedric Hamiltonian of $H_{\kappa=0}$.

Gauge-invariant Hamiltonians such as $H_{dual3}$ have been
employed in the past as models for {\it open} surfaces \cite{10a}, since the use of edge spins allows spin clusters to have free edges and seams \cite{3}. Indeed, the archetypal lattice gauge-matter theory, the $3d$ $\mathbb{Z}_2$ gauge-Higgs model \cite{10b}
\begin{equation}
H= - \beta_1 \sum_{\langle ij \rangle}^{ }(\sigma_{i}U_{ij}\sigma_{j}) -
\beta_4 \sum_{[ijkl]}^{ }U_{ij}U_{jk}U_{kl}U_{li} \, ,
\label{3DZ2}
\end{equation}
has itself been used in such a context \cite{11}. In this paper the gauge spins $U^{1,2,3}$ of $H_{dual3}$ are non-dynamical, so it would be an interesting extension of the current investigations to  include a 
pure gauge term in the Hamiltonian in the manner
of equ.~(\ref{3DZ2}) to observe the interplay between the anisotropic matter
couplings and the gauge spins. 

Considering the original plaquette Hamiltonian  $H_{\kappa=0}$
it is also possible to write down a gauge-Ising variant 
that allows open surfaces, which takes the form \cite{12}
\begin{eqnarray}
H &=& - \beta_2 \sum_{[ijkl]}^{ }
\left[ (\sigma_{i} U_{ij}\sigma_{j})(\sigma_{k}U_{kl} \sigma_{l}) +(\sigma_{i}U_{il} \sigma_{l})(\sigma_{j}U_{jk} \sigma_{k}) \right] \nonumber \\
&-& \beta_4 \sum_{[ijkl]}^{ }(U_{ik}U_{jk}U_{kl}U_{li})
\label{GG}
\end{eqnarray}
where the matter couplings are dimer sums over the opposite edges of plaquettes. In this case the anisotropy seen in the dual Hamiltonians is not present.

We close by repeating our observation in \cite{5b} regarding $H_{dual2}$ and $H_{dual1}$: it is a curious feature
of these dual Hamiltonians, and as we have seen in this paper $H_{dual3}$ also, that they are all anisotropic in spite of being dual
to an isotropic Hamiltonian. The flip symmetries that are a direct consequence of this, and the gauge symmetry
manifest in $H_{dual3}$ and inherited by $H_{dual2}$, play an important role in determining the properties of the models and how the different Hamiltonians are related.

The coupling between the spin types and spatial directions in the dual gonihedric models is reminiscent of 
another class of anisotropically coupled Hamiltonians, the compass models, which exist in both classical and quantum forms \cite{14}. 
In $3d$ for instance, the quantum compass Hamiltonian is of the form \cite{15}
\begin{equation}
 H_{compass} = - J_x \sum_{\langle ij \rangle} \sigma^x_{i}  \sigma^x_{j} 
- J_y  \sum_{\langle ik \rangle } \sigma^y_{i}  \sigma^y_{k} 
- J_z  \sum_{ \langle jk \rangle} \sigma^z_{j} \sigma^z_{k} 	\, .
\end{equation}
where the sums are again one dimensional and the $\sigma$  are now  Pauli matrices. These too
are know to display numerous interesting symmetries  \cite{16}, and have been found to have
strong finite-size effects with periodic boundary conditions \cite{17}. The parallels between these
models and the gonihedric Hamiltonians merit further investigation, as do the potential numerical
pitfalls involved in simulations of both.  
 
\section{Acknowledgements}
The work of R. P. K. C. M. Ranasinghe was supported by a Commonwealth Academic Fellowship {\bf LKCF-2010-11}.



\bigskip
\bigskip
\bigskip
\bigskip
\bigskip
\bigskip
\bigskip




\begin{thebibliography}{}

\bibitem{1} R.V. Ambartzumian, G.S. Sukiasian, G. K. Savvidy
            and K.G. Savvidy, Phys. Lett. {\bf B275} (1992) 99.\\
            G. K. Savvidy and K.G. Savvidy, Int. J. Mod. Phys.
            {\bf A8} (1993) 3393.\\
            G. K. Savvidy and K.G. Savvidy, Mod. Phys. Lett.
            {\bf A8} (1993) 2963.\\
            J. Ambj\o rn, G.K. Savvidy and K.G. Savvidy, Nucl.Phys. {\bf B486} (1997) 390.



\bibitem{2} G. K. Savvidy and F.J. Wegner, Nucl. Phys. {\bf B413} (1994) 605.\\
            G. K. Savvidy and K.G. Savvidy, Phys. Lett. {\bf B324} (1994) 72.\\ 
            G. K. Savvidy and K.G. Savvidy, Phys. Lett.{\bf B337} (1994) 333.\\
            G.K.Bathas, E.Floratos, G.K.Savvidy and K.G.Savvidy, Mod. Phys. Lett. {\bf A10} (1995) 2695.\\
            G. K. Savvidy and K.G. Savvidy, Mod. Phys. Lett. {\bf A11} (1996) 1379.\\
            G. Koutsoumbas, G. K. Savvidy and K. G. Savvidy, Phys.Lett. {\bf B410} (1997) 241.\\
            J.Ambj\o rn, G.Koutsoumbas, G.K.Savvidy, Europhys.Lett. {\bf 46} (1999) 319.\\
            G.Koutsoumbas and G.K.Savvidy, Mod.Phys.Lett. {\bf A17} (2002) 751.\\
            D. Johnston and R.K.P.C. Malmini, Phys. Lett. {\bf B378} (1996) 87.\\
            M. Baig, D. Espriu, D. Johnston and R.K.P.C. Malmini, J. Phys. {\bf A30} (1997) 405.\\
            M. Baig, D. Espriu, D. Johnston and R.K.P.C. Malmini, J. Phys. {\bf A30} (1997)  7695.

\bibitem{3}  A. Cappi, P. Colangelo, G. Gonella and A. Maritan, Nucl. Phys. {\bf B370} (1992) 659.

\bibitem{4} A. Lipowski J. Phys. {\bf A30} (1997) 7365.\\
             A. Lipowski and D. Johnston, J. Phys. {\bf A33} (2000) 4451.\\
             A.Lipowski and D.Johnston, Phys.Rev. {\bf E61} (2000) 6375.\\
             A. Lipowski, D. Johnston and D. Espriu, Phys Rev. {\bf E62}  (2000) 3404.
             M. Swift, H. Bokil, R. Travasso and A. Bray, Phys. Rev. {\bf B62} (2000) 11494.\\
             A. Cavagna, I. Giardina and T. S. Grigera, Europhys.Lett. {\bf 61} (2003) 74; J. Chem. Phys. {\bf 118} (2003) 6974.\\
             S. Davatolhagh, D. Dariush and L. Separdar, Phys Rev. {\bf E81} (2010) 031501.

\bibitem{5}  D. Johnston and R.K.P.C.M. Ranasinghe, J. Phys. {\bf A44} (2011) 295004.

\bibitem{5b}  D. Johnston and R.K.P.C.M. Ranasinghe, ``Another Dual gonihedric 3D Ising Model'', {\tt [arXiv:1106.0325 ]}.

\bibitem{6} G. K. Savvidy, K.G. Savvidy and P.G. Savvidy, Phys.Lett. {\bf A221} (1996) 233.

\bibitem{AT} J. Ashkin and E. Teller, Phys. Rev. {\bf 64} (1943) 178.


\bibitem{6b} G. K. Savvidy, K.G. Savvidy and F.J. Wegner, Nucl. Phys. {\bf B443}
            (1995) 565.
            

\bibitem{6c}  F.J. Wegner,J. Math. Phys. {\bf 12} (1971) 2259.

\bibitem{7} 
I Syozi, in: C. Domb. M.S. Green (Eds.) Phase Transitions
and Critical Phenomena, {\it vol 1}, Academic Press, New York (1972) 269.

\bibitem{10} R. Pietig and F. Wegner, Nucl.Phys. {\bf B466} (1996) 513.\\
              R. Pietig and F. Wegner, Nucl.Phys. {\bf B525} (1998) 549.     

\bibitem{10a} A. Cappi, P. Colangelo, G. Gonella and A. Maritan, Nucl. Phys. {\bf
B370} (1992) 659.

\bibitem{10b} M. Creutz, Phys. Rev. {\bf D21} (1980) 1006.\\
              G. Bhanot and M. Creutz, Phys. Rev. {\bf D21} (1980) 2892.\\
              F. Gliozzi and A. Rago, Phys.Rev. {\bf D66} (2002) 074511. 

\bibitem{11} D. Huse and S. Leibler, Phys. Rev. lett. {\bf 66} (1991) 437.

\bibitem{12} R.K.P.C.M. Ranasinghe, J.Natn.Sci.Foundation Sri Lanka {\bf 36} (2008) 299.			

\bibitem{14} K. I. Kugel and D. I. Khomskii, Sov. Phys. Usp. {\bf 25} (1982) 231.\\
D. I. Khomskii and M. V. Mostovoy, J. Phys. {\bf A36} (2003)  9197.\\
M. V. Mostovoy and D. I. Khomskii, Phys. Rev. Lett. {\bf 92} (2004) 167201.\\
J. van der Brink, New J. Phys. {\bf 6} (2004) 201.\\
G. Jackeli and G. Khaliullin, Phys. Rev. Lett. {\bf 102} (2009) 017205. 

\bibitem{15} J. Oitmaa and C. J. Hamer,  Phys. Rev. {\bf B83} (2011) 094437.

\bibitem{16} Z. Nussinov and E. Fradkin, Phys. Rev. {\bf B71} (2005) 195120. 

\bibitem{17} S. Wenzel and W. Janke, Phys. Rev. {\bf B78} (2008) 064402.\\
S. Wenzel, W. Janke, and A. Läuchli, Phys. Rev. {\bf E81} (2010) 066702. 


\end{thebibliography}
\end{document}